\documentclass[fleqn,twocolumn]{openjournal}

\usepackage[T1]{fontenc}
\usepackage{ae,aecompl}
\usepackage{graphicx}   
\usepackage{hyperref}
\usepackage{lipsum} 
\usepackage[sort&compress]{natbib}
\usepackage{ulem}
\usepackage{amsmath}    
\usepackage{amssymb}    
\usepackage{natbib}
\usepackage{color}      
\definecolor{dgreen}{rgb}{0,0.5,0}
\definecolor{dp}{rgb}{0.5,0,0.5}
\definecolor{magen}{rgb}{0.79,0.08,0.48}
\definecolor{darkred}{rgb}{0.65,0.06,0.37}
\usepackage{booktabs}
\usepackage{multirow}
\usepackage{makecell}

\begin{document}

\nocite{*}

\title[FOOD I]{FOOD I: A New Division Scheme For The Stelliferous Era}

\author{Stephen M. Wilkins$^{1}$, Sophie L. Newman$^{2}$, Will J. Roper$^{1}$}
\affil{$^{1}$Astronomy Centre, University of Sussex, Falmer, Brighton BN1 9QH, United Kingdom}
\affil{$^{2}$Institute of Cosmology and Gravitation, Dennis Sciama Building, Burnaby Road, Portsmouth, PO1 3FX}
\email{s.wilkins@sussex.ac.uk}


\begin{abstract}
In recent years the James Webb Space Telescope has enabled the frontier of observational galaxy formation to push to ever higher redshift, deep within cosmic dawn. However, what is high-redshift, and when was cosmic dawn? While widely used, these terms (as well as many other confusing terms) are not consistently defined in the literature; this both hampers effective communication but also impedes our ability to precisely characterize and understand the phenomena under investigation. In this article we seek to address this issue of utmost importance. We begin by definitively defining terms such as ``high-redshift'', ``cosmic dawn'', etc.  However, despite the rigorous definitions for them we present, both the adjective-based redshift and diurnal marker (time-of-day) division schemes suffer from issues including not being sufficiently granular, angering cosmologists, being arbitrary, and having a geocentric bias. To overcome these we introduce the \textit{redshiFt epOchs fOr everyboDy} (FOOD) framework, a revolutionary new division scheme based on eating occasions, i.e. meals. 
\end{abstract}

\maketitle

\section{Introduction}

The exploration of the high-redshift Universe is key to understanding galaxy formation, the Epoch of Reionization (EoR), and cosmology. With the advent of cutting-edge telescopes like the James Webb Space Telescope (JWST), galaxies and AGN have been observed and spectroscopically confirmed to higher redshifts than ever (e.g. \citealt{bunker}, \citealt{naidu}). However, amid these groundbreaking discoveries lies a fundamental challenge: nomenclature! Or more specifically, the lack of standardized terminology and classification systems.

Terms like ``high-redshift'' and ``cosmic dawn''\footnote{Don't even get us started on the use of ``first light'' - this one really draws the ire of cosmologists. Taken at face value this term should almost certainly denote a regime at $z=\infty$ since there have always been photons, but it is awash with abuse \citep[e.g][]{FLARES-I, FLARES-II, FLARES-III, FLARES-IV, FLARES-V, FLARES-VI, FLARES-VII, FLARES-VIII, FLARES-IX, FLARES-X, FLARES-XI, FLARES-XII, FLARES_XIII, FLARES-XIV}.} are pervasive in cosmological discourse (e.g. \citealt{supernovae}, \citealt{2010MNRAS.408.2115M}, \citealt{2023ApJ...959L..14N}, \citealt{2024MNRAS.527.9977S}), yet their definitions vary widely across different studies and disciplines. This inconsistency not only hampers effective communication but also impedes our ability to precisely characterize and understand the phenomena under investigation\footnote{Although we do acknowledge that the freedom to throw around these buzzwords in grant proposals has benefited many within the field, ourselves included.}. 
Moreover, existing division schemes based on redshift or time intervals suffer from limitations such as insufficient granularity and geocentric biases, hindering their applicability and universality.

In this article, we embark on a journey to rectify these shortcomings. First, we clearly define the divisions of both an adjective-based redshift (Section \ref{sec:redshift}) and diurnal marker (time-of-day) based scheme (Section \ref{sec:time_of_day}). Second, in Section \ref{sec:culinary} we introduce the redshiFt epOchs fOr everyboDy (FOOD) nomenclature—a revolutionary framework for categorizing high-redshift events based on a universally relatable concept: meals. 
The universality of meals means translating from one division scheme to another within the FOOD framework should be trivial to anyone in need of food, a fact we will demonstrate with a parallel scheme at the end of this work. 
By aligning cosmic epochs with common eating occasions, we aim to provide a more intuitive and inclusive system that transcends disciplinary boundaries.

\section{Adjective-based redshift division scheme}\label{sec:redshift}

We begin by tackling a question that has troubled astrophysicists for several decades - what does ``high-redshift'' even mean? 

In its simplest terms, high-redshift galaxies are galaxies that are observed at great distances from Earth, corresponding to an era when the Universe was much younger. However, we need to know how far is far enough (or how early is early enough) for an object to truly count as ``high-redshift''. Moreover, it remains an open question whether there are objects at ``higher-redshift'' than ``high-redshift'' and if they're not at ``high-redshift'' what adjective should be used to describe them? Once and for all we settle this question by defining an adjective-based division scheme that both extends to the highest redshifts imaginable and offers sufficient high-redshift granularity to enable exciting-sounding grant proposal titles.







\begin{figure}
    \centering
    \includegraphics[width=1\columnwidth]{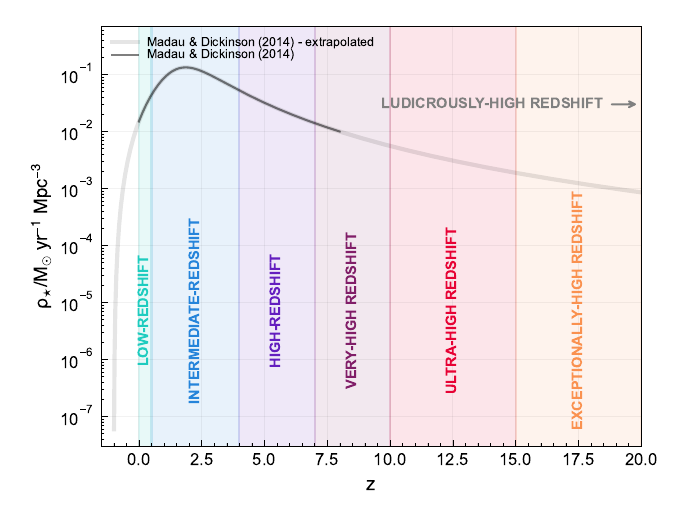}
    \caption{A visual representation of our proposed "redshift" division scheme. The cosmic star formation history of \citet{MD14}, including an extrapolation at $z<0$ and $z>8$ is shown for reference. }
    \label{fig:redshift}
\end{figure}

We begin by defining the three well established adjective divisions: ``low'', ``intermediate'', and ``high'' redshift. First, in common with previous authors we define ``low-redshift'' to correspond to $z<0.5$. We then choose to define intermediate redshift such that the peak of the cosmic star formation history \citep[see e.g.][]{MD14, Wilkins2019}, which occurs at $z\approx 2.0$, lies approximately in the centre of the band, i.e. $0.5<z\le 4$. While the lower-band of the high-redshift sub-division is established by the intermediate redshift division the upper-bound is more problematic. While in principle this could be unbound this provides too little granuality to excite grant and article reviewers. Consequently we choose to define the upper-limit at $z=7$; roughly corresponding to the redshift at which the Lyman-break is shifted out of the visible and into the near-infrared. This also corresponds, approximately, to the highest redshifts observationally accessible (for individual galaxies) prior to installation of wide field camera three (WFC3) on the Hubble Space Telescope in 2009 \citep[e.g.][]{Wilkins2010}. Next, we define a division extending to $z=10$ which we call "very-high" redshift, an era that was terribly exciting back before JWST. Here the maximum limit, in addition to being a nice round number, was also the approximate limit of Hubble/WFC3 observations, with the notable exception of GN-z11 at $z=10.6$ \citep{Oesch2016}. With the arrival of JWST we are now able to identify, and crucially confirm, galaxies at $z>10$, with the current confirmed record holder at $z=13.2$ \citep{CurtisLake2023}. As new deeper observations are obtained it is inevitable that this frontier will extend to $z=15$, and potentially surpass it. Since $z=15$ is a nice number we define this as the upper-limit of the "ultra-high" redshift division. As JWST has the technological capability to identify sources at $z>15$ we refer to that division as "exceptionally" high-redshift. To keep cosmologists happy we also define "ludicrously" high-redshift to encompass recombination and the Universe's preceding history. This way they can stop confusing astronomers by referring to high-redshift while meaning a regime when the Universe has existed for mere minutes. These divisions are defined in Table \ref{tab:redshift} and visually compared to the \citet{MD14} cosmic star formation history in Figure \ref{fig:redshift}.

\begin{table}
\centering
\caption{Division scheme based on redshift. \label{tab:redshift}}
\begin{tabular}{l|rcl}
  \textbf{low-redshift}   &   & $z$ & $\le 0.5$\\
  \textbf{intermediate-redshift}    & $0.5<$ & $z$ & $\le 4$\\ 
  \textbf{high-redshift}   & $4<$ & $z$ & $\le 7$\\
  \textbf{very-high redshift}    & $7<$ & $z$ & $\le 10$\\
  \textbf{ultra-high redshift}    & $10<$ & $z$ & $\le 15$\\
  \textbf{exceptionally-high redshift}    & & $z$ & $>15$\\
  \textbf{ludicrously-high redshift}      & & $z$ & $\gtrsim1000$
\end{tabular}
\end{table}

\section{Diurnal marker division scheme}\label{sec:time_of_day}

\begin{figure}
    \centering
    \includegraphics[width=1\columnwidth]{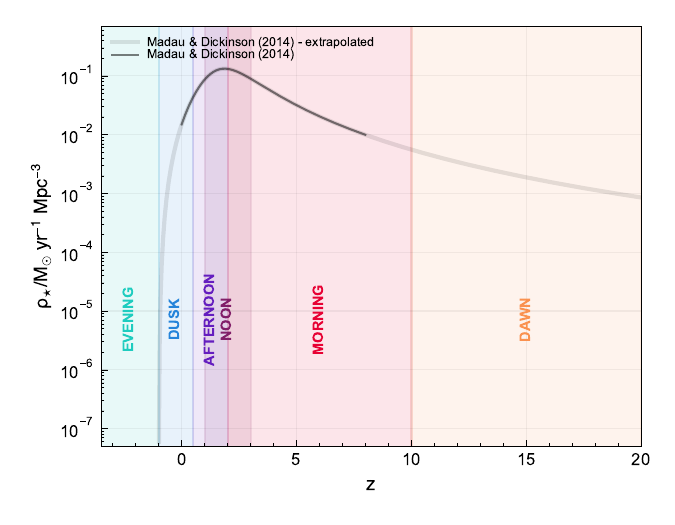}
    \caption{A visual representation of our proposed diurnal marker "redshift" division scheme. The cosmic star formation history of \citet{MD14}, including an extrapolation at $z<0$ and $z>8$ is shown for reference. }
    \label{fig:times_of_day}
\end{figure}

Adjective-based redshift division schemes have always been controversial since one researcher's ``high-redshift'' is another's ``low-redshift'', particularly if they work on the cosmic microwave background ($z\approx 1000$) or inflation ($z\approx 10^{25-30}$). To avoid provoking the wrath of early Universe cosmologists we instead propose a division scheme based on times of the day (here on Earth), i.e. "dawn", "morning", "noon", etc. Such a scheme is not an entirely new innovation as "cosmic dawn"\footnote{While terms such as "cosmic dawn" inherently sound better, terms like "stellar" or "galaxy dawn" are more accurate and less likely to incur the ire of cosmologists. 
} is already in regular use. Our innovation is to precisely define "cosmic dawn" and extend to include "cosmic morning", "cosmic noon", "cosmic afternoon", "cosmic dusk", and "cosmic evening". Here "cosmic noon" is chosen to align with the peak of the cosmic star formation history ($z=2$), but to also encompass a range $1<z\le 3$ to account for uncertainty in the exact location of the peak. Either side of the peak we define "cosmic afternoon" and "cosmic morning" both of which partially overlap with "cosmic noon". For the epoch preceding "cosmic morning" we introduce "cosmic dawn" extending beyond $z>10$. We define "cosmic dusk" as the current epoch during which the cosmic star formation history is rapidly decreasing. For completeness we also define "cosmic evening" ($z<-1.0$) to coincide with the epoch where the star formation rate density has fallen to approximately $1\%$ of its peak. These divisions are defined in Table \ref{tab:times_of_day} and visually compared to the \citet{MD14} cosmic star formation history in Figure \ref{fig:times_of_day}.

\begin{table}
\centering
\caption{Division scheme based on times of the day. \label{tab:times_of_day}}
\begin{tabular}{l|rcl}
  \textbf{Dawn}   &  $10<$ & $z$ & $\le 20$\\
  \textbf{Morning}   & $2<$ & $z$ & $\le 10$\\
  \textbf{Noon}   & $1<$ & $z$ & $\le 3$\\
  \textbf{Afternoon}   &$0.5<$  & $z$ & $\le 2$\\
  \textbf{Dusk}   &$-1.0<$  & $z$ & $\le 0.5$\\
  \textbf{Evening}   & & $z$ & $\le -1.0$\\
\end{tabular}
\vspace{10pt} 
\end{table}

\section{A meal time based division scheme}\label{sec:culinary}

\begin{figure}
    \centering
    \includegraphics[width=1\columnwidth]{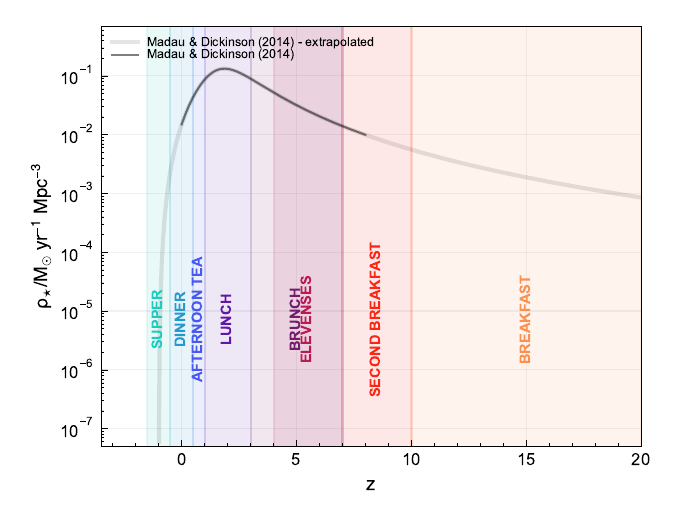}
    \caption{A visual representation of the FOOD "redshift" division scheme. The cosmic star formation history of \citet{MD14}, including an extrapolation at $z<0$ and $z>8$ is shown for reference. }
    \label{fig:culinary1}
\end{figure}

Like the adjective-based redshift division scheme, a time of day based scheme has several disadvantages. Chief here is that it is not universal and only applies to planets that are not tidally locked to their parent star. To overcome these limitations we introduce a new scheme based on popular eating occasions (which are sure to be universal not just for Humanity). This scheme is presented in Table \ref{tab:culinary1} and visually compared to the \citet{MD14} cosmic star formation history in Figure \ref{fig:culinary1}. Compared to the other schemes this offers the advantage of increased granularity and tastiness. 

In designing this scheme we begin with the three (almost) universally acknowledged meals: breakfast, a midday meal (lunch), and an early evening meal (dinner). However, we augment these meals with several others including the more common "brunch", "afternoon tea", and "supper", and for further high-redshift granularity include the most important of meals "second breakfast" and "elevenses". 

Here we acknowledge those deeply inspired by the academic works of Tolkien (e.g. \citealt{LOTR1}, \citealt{LOTR2}, \citealt{LOTR3}, \citealt{hobbit}, \citealt{tom}) set in the region of Middle-earth, where meals play a central role in the daily lives of its inhabitants, reflecting both cultural traditions and the rhythm of daily routines. Among the most beloved and whimsically named meals are "elevenses" and "second breakfast," which offer delightful glimpses into the culinary customs of the Hobbits, particularly those of the Shire. Elevenses, occurring around mid-morning, is a time for a light repast to tide one over until the main meal of the day. Typically consisting of freshly baked bread, butter, cheese, and perhaps a slice of fruitcake or a sweet treat, elevenses encapsulates the Hobbit penchant for simple yet satisfying fare. Second breakfast, as the name suggests, follows shortly after, providing a hearty refueling to sustain the industrious Hobbit through the morning hours. Including these Tolkien-inspired time divisions provides a way to be more specific about what redshift domain is being referred to, as well as invoking warm and happy feelings to the academic community which will result in more productivity.

In terms of the exact limits of each division we define "cosmic lunch" to coincide with "cosmic noon". Beyond this we simply make things up\footnote{A long standing tradition in Astronomy still going strong in this age of next generation telescopes.}.

\begin{table}
\centering
\caption{Division scheme based on meal times. \label{tab:culinary1}}
\begin{tabular}{l|rcl}
  \textbf{Breakfast}   &  $10<$ & $z$ & $\le 20$\\
  \textbf{Second breakfast}  & $7<$  & $z$ & $\le 10$\\
  \textbf{Elevenses}   & $4<$  & $z$ & $\le 7$\\
  \textbf{Brunch}   & $3<$  & $z$ & $\le 7$\\
  \textbf{Lunch}   & $1<$  & $z$ & $\le 3$\\
  \textbf{Afternoon tea}   & $0.5<$  & $z$ & $\le 1$\\
  \textbf{Dinner}  & $-0.5<$  & $z$ & $\le 0.5$\\
  \textbf{Supper}  &  $-1.5<$ & $z$ & $\le -0.5$\\
\end{tabular}
\vspace{10pt} 
\end{table}

\subsection{An alternative scheme for (some) users from the North of England and other areas}

To demonstrate the universality of the FOOD framework, we modify the divisions presented in Table \ref{fig:culinary1} to be inclusive of those from Northern England\footnote{Though there is no universal agreement, largely because everyone else is wrong, "The North" corresponds to the former borders of Anglian Northumbria, the Anglo-Scandinavian Kingdom of Jorvik and the Brythontic Celtic Hen Ogledd kingdoms. The region immediately south of the North, is the Midlands, and south of the midlands is the South (including London).}\footnote{The third author would like to argue that "The North" begins at the northern boundary of Greater London. Consistent with the region north of which the English accent takes on a life of its own.}\footnote{The first author would like to point out to the third author that there are plenty of lively English accents south of north London, unless one is willing to ignore the West Country.}\footnote{The third author acknowledges the first author's point but chooses to give the West Country a pass. Their accent's endearing lilt is reason enough, but one could also argue for a further division placing them in "The West" along with the Cornish.}. 

As with much of British culture, the origins of "dinner time" are buried deep within extreme class divisions (if we're honest this is a refreshing change from the colonial origins of the rest of British culture). 

\begin{figure}
    \centering
    \includegraphics[width=1\columnwidth]{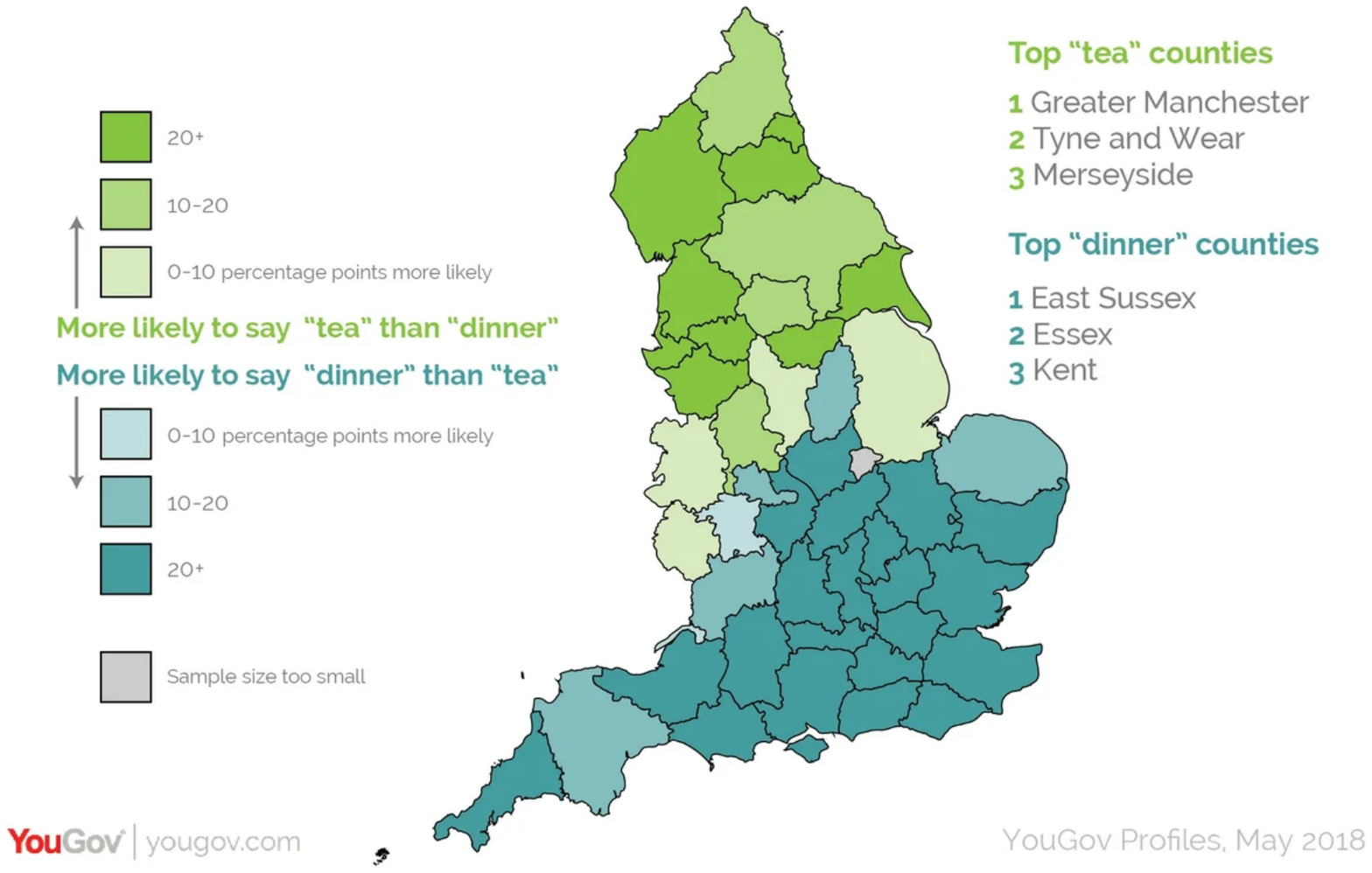}
    \caption{A map showing the relative abundance of tea time vs dinner time in England taken from \cite{dvt_map_yougov}.}
    \label{fig:dvt_map}
\end{figure}

Historically, the main meal of the day was denoted "dinner", however, dinner meant different things for different classes of people. For children and manual workers, dinner was taken at midday; in the early evening for office workers; and in the late evening by the wealthier elements of society. 
During the latter half of the 20$^\mathrm{th}$ century, there has been a cultural shift towards everyone having the main meal in the late evening (see \citealt{dinner} for more details). Thus dinner has come to mean the early evening meal across most parts of the English-speaking world. However, in some parts of the world, particularly in the North (of England), dinner has remained the name for the midday meal amongst a sizeable fraction of the population. In these areas, the early evening meal is often referred to as "tea". Indeed, food supplier Geest surveyed 1000 British people and found that of those in Northern England 68\% called the main evening meal "tea", whereas only 5\% of those in London followed the same custom
. We show the rough modern-day distribution of "tea time" vs "dinner time" in Figure \ref{fig:dvt_map} (courtesy of \cite{dvt_map_yougov}). To accommodate this cultural difference we introduce a parallel division scheme where "lunch" is replaced by "dinner", "dinner" by "tea", and "afternoon tea" by "afternoon snack". 

\begin{table}
\caption{Alternative to the scheme presented in Table \ref{tab:culinary1}. \label{tab:culinary2}}
\centering
\begin{tabular}{l|rcl}
  \textbf{Breakfast}   &  $10<$ & $z$ & $\le 20$\\
  \textbf{Second breakfast}  & $7<$  & $z$ & $\le 10$\\
  \textbf{Elevenses}   & $4<$  & $z$ & $\le 7$\\
  \textbf{Brunch}   & $3<$  & $z$ & $\le 7$\\
  \textbf{Dinner}   & $1<$  & $z$ & $\le 3$\\
  \textbf{Afternoon snack}   & $0.5<$  & $z$ & $\le 1$\\
  \textbf{Tea}  & $-0.5<$  & $z$ & $\le 0.5$\\
  \textbf{Supper}  &  $-1.5<$ & $z$ & $\le -0.5$\\
\end{tabular}
\vspace{10pt} 
\end{table}

\begin{figure}
    \centering
    \includegraphics[width=1\columnwidth]{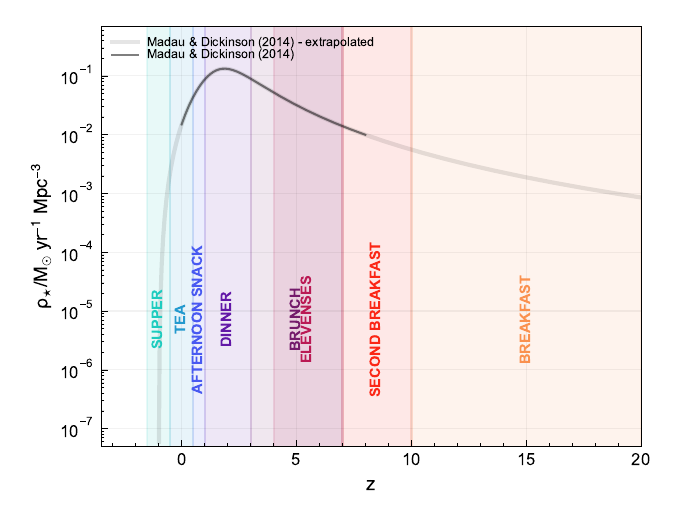}
    \caption{A visual representation of the parallel FOOD "redshift" division scheme. The cosmic star formation history of \citet{MD14}, including an extrapolation at $z<0$ and $z>8$ is shown for reference. }
    \label{fig:northern_redshift}
\end{figure}

\section{Conclusions}
\label{sec:conclusions}

We have settled the age-old debate of what redshifts count as high-redshift and when cosmic dawn occurred by unilaterally defining these terms for the community. We predict this will save countless hours of debate. Importantly we have also defined a term for cosmologists so they no longer need to improperly use Astronomer's nomenclature. 

In addition, we propose the redshiFt epOchs fOr everyboDy (FOOD) framework, a new division scheme for the stelliferous era using a meal time-based scheme. While an obvious improvement, this scheme does have the disadvantage of requiring a second parallel scheme to account for cultural differences in the naming of meals. Some may say we have only further deepened confusion in including this modified scheme. However, the authors are sure those familiar with the portion of the British population unable to agree on the name of a bread roll should have a method by which to subtly identify themselves so the rest of us can avoid them\footnote{See \href{https://yougov.co.uk/consumer/articles/21204-cobs-buns-baps-or-barm-cakes-what-do-people-call-b}{here}.}.

We hope this work will prove useful to the wider community. We invite anyone to produce their own parallel schemes for meal times within their own culture. These parallel schemes, rooted in the universality of FOOD, will only serve to enrich our understanding and communication.

\begin{acknowledgments}
\section*{Acknowledgements}
The authors would like to acknowledge the contribution of many in the Astronomy Centre at the University of Sussex for their insightful discussions on the translation of meals into cosmic epochs. WJR would like to thank his collaborators for providing enough work (and in the case of Peter Thomas a healthy dose of cynicism) that he was unable to further the COWSHED project \citep{COWSHEDI, COWSHEDII} in 2024. 
\end{acknowledgments}


\vspace{5cm}

\newpage 
\bibliography{bib.bib}
\end{document}